# Breaking the thermally induced write error in heat assisted recording by using low and high Tc materials

D. Suess,

*Vienna University of Technology, Wiedner Hauptstrasse 8-10, 1040 Vienna, Austria.*

T. Schrefl

*University of Applied Sciences, Matthias Corvinus-Straße 15, 3100 St. Poelten, Austria*

**Abstact:** Heat assisted recording is believed as a key technology in order to further increase the areal density of magnetic recording. In the work of Richter et al. [Richter et al. J. Appl. Phys. 111, 033909 (2012)] it is stated that storage densities will be limited to 15 to 20 Tbit/in² due to thermally induced write errors. In this letter we propose a composite structure consisting of two materials with different Curie temperatures. A hard magnetic layer is on top of a high $T_c$ soft magnetic layer. In this composite material the thermal write error is negligible up to areal densities of 50 – 100 Tbit/in². It is shown that the effective thermal field gradient, which is reduced in this composite structure, is not relevant for a possible increase of the transition jitter. The transition jitter is dominated by a small distribution of the Curie temperature. The smallest jitter is obtained for the composite structure.

In magnetic recording the areal density is believed to be limited by the recording trilemma [1]. In order to break the recording trilemma various scheme are proposed such as bit patterned media [2], inhomogeneous reversal schemes (ECC [3], exchange spring and graded media [4]) and heat assisted recording [5,6]. ECC media and exchange spring media are widely discussed in academia and industry in order to gain beneficial effects in heat assisted recording. In Ref it is shown that ECC media with a soft magnetic layer on top of a hard magnetic layer reduce anisotropy fluctuation in heat assisted recording.

Recently, the recording trilemma is extended to be a recording quadrilemma [7]. Richter et al. have shown that thermally induced recording errors occur due to thermodynamical processes during the recording process. These errors are negligible for recording at room temperature. However, for the future recording concept of heat assisted recording these errors are believed to be the limiting factor [8].

In the work of Richter et al. FePt is investigated as heat assisted recording media. The material properties are assumed to scale as function of temperature as follows:

$$J_s(T) = J_s(T_0)\sqrt{\left(1-(T/T_c)^2\right)} \qquad (1)$$

$$K_1(T) = K_1(T_0)(M_s(T)/M_s(T_0))^2 \qquad (2)$$

The temperature dependence of these properties is shown in Fig. 1. In order to record the ultra- high coercive material FePt the temperature has to be as high as $T_{wr} = 746$ K during the recording process in order to lower the coercive field to values around 1.6 T. Here we have assumed that the recording head can generate a field of 1.6 T. This temperature is just below the Curie temperature of $T_c = 750$ K. As a consequence of the high temperature not only the anisotropy decreases but also the saturation polarization $J_s(T_{wr}) = \mu_0 M_s(T_{wr})$ decays significantly ($M_s$ is the saturation magnetization). Due to the small saturation polarization at this high temperature thermally induced write errors occur. In the one grain per bit limit the BER (bit error rate) can be expressed as [7],

$$BER = \exp\left(-\frac{2J_s(T_{wr})VH_{wr}}{k_B T_{wr}}\right), \qquad (3)$$

,where $T_{wr}$ is the temperature during the write process, $H_{wr}$ the maximum write field, $k_B$ the Boltzman constant, $V$ the particle volume which is equal to the bit volume and BER denotes the bit error rate. In Ref. (7) it is assumed that the relation of the areal density (AD) and the particle volume $V$ can be approximated as

$$AD = \alpha V^{-2/3}, \qquad (4)$$

where $\alpha$ is the areal packing fraction of the storage islands, which is assumed to be $\alpha = 0.5$. Substituting Eq. (4) into Eq. (3), the areal density can be expressed as function of the BER [7]:

$$AD = \alpha \left( \frac{2 J_s(T_{wr}) H_{wr}}{k_B T_{wr} \ln(1/BER)} \right)^{2/3}, \qquad (5)$$

The switching field as function of temperature is estimated by the Stoner – Wohlfarth theory using:

$$\mu_0 H_s(T) = \mu_0 \frac{2 K_1(T)}{J_s(T)} \qquad (6)$$

The temperature dependence of the anisotropy constant $K_1(T)$ and the saturation polarization $J_s(T)$ for FePt is shown in Fig. 1. Fig. 3 shows the switching field according to Eq. (4) as function of temperature. At the temperature of $T_{wr} = 746$ K, where the switching field is sufficient small (1.6T) that it can be reversed with the write head, the magnetic polarization is $J_s = \mu_0 M_s(T_{wr}) = 0.17 T$. This small saturation polarization leads to the above mentioned thermally induced write errors. The areal density (AD) as function of BER is shown in Fig. 4 for the FePt media using Eq. (5).

In this letter we present two simple media grain designs which allow overcoming the thermally induced write error. The media design consists of a composition of two materials. One material having a small Curie temperature and the other material having a large Curie temperature.

As an example we investigate the recently fabricated FePt / Fe exchange spring media [9,10]. The Curie temperature of Fe is $T_c = 1040$ K [11]. Due to the fact the heating of the media is restricted to the surface of the media[12], we assume that the high coercive with low $T_c$ material is located next to the pole tip and the softer high $T_c$ material below it. This arrangement of soft/hard layers is opposite to commonly employed exchange spring and ECC media. The temperature dependence of the saturation polarization (Fig. 1) is extracted from Ref. ([11]). The magnetic anisotropy of Fe can be neglected. In the following we assumed media (a: Fe/FePt – single domain) having a column length smaller than the exchange length of the hard magnetic material $l = \sqrt{A/K_1}$. As a consequence the reversal process occurs almost coherent in the soft magnetic and hard magnetic region [13–15]. For this coherent rotation the material parameters which influence the reversal process can be averaged over the entire media grain. In the following we assume a fraction of 50% FePt and 50% Fe, respectively. The average saturation polarization and magnetic anisotropy are shown in Fig. 1. Due to the fact that the Curie temperature of Fe is much larger than that of FePt the average saturation polarization is still large, close to the Curie temperature of FePt. The switching field as function of temperature calculated with Eq. (6) is shown in Fig. 2. The switching field is 1.6 T at a temperature of $T_{wr}$ = 683 K. The average saturation polarization at this temperature is as high as $J_r$ = 1.2 T. Since the thermally induced write error decay exponentially with the magnetic polarization (Eq. (3)) small write errors occur allowing for low BER. For a particle size which correspond to an $AD$ = 30Tb/in² the $BER$=1.1 x $10^{-5}$. The areal density as function of BER is shown in Fig. 4 allowing for AD up to 50 Tbit/inch², as calculated according to Eq. (5)

As a second media design (b: Fe/FePt – exchange spring) the column length of the recording grain is increased above the exchange length of the hard magnetic material. As a consequence the reversal process occurs via a formation of a domain wall. The switching field can be calculated as given by Kronmüller and Goll [16] using

$$\mu_0 H_s = \mu_0 \frac{2K_{hard}}{J_{s,hard}} \frac{1-\varepsilon_K \varepsilon_A}{\left(1+\sqrt{\varepsilon_J \varepsilon_A}\right)^2} . \tag{7}$$

The ratio of the material parameters of the soft layer to the hard layer are given by the ratio of the anisotropy constants, $\varepsilon_K = K_{soft}/K_{hard}$, the ratio of the exchange constants, $\varepsilon_A = A_{soft}/A_{hard}$, and the ratio of the magnetic polarizations, $\varepsilon_J = J_{s,soft}/J_{s,hard}$. For the temperature dependence of the exchange constant we assume [17],

$$A(T) = A(T_0)\left(M_s(T)/M_s(T_0)\right)^{1.7} . \tag{8}$$

The switching field as function of temperature is shown in Fig. 3. Here, we have used the temperature scaling of the material properties in each material phase according to the equations given above. The switching field is calculated using Eq. (7). Almost no thermal write assist is required in order to switch the exchange spring media grain. The switching field is 1.6 T at a temperature of $T_{wr}$ = 318 K. The saturation magnetization at this temperature is as high as $J_r = 1.75$ T. The thermally induced write error allows for areal densities above 100 Tbit/inch².

Finally, we investigate the position jitter of the three different media designs. Two contributions of transition jitter are investigated. (i) jitter due to variation of the anisotropy constant $\sigma_{K1}$ (ii) jitter due to variations of the Curie temperature $\sigma_{Tc}$.

For the calculation of the jitter contribution (i) the thermal effective head field gradient is evaluated as,

$$\mu_0 \frac{\partial H_s}{\partial x} = \mu_0 \frac{\partial H_s}{\partial T} \frac{\partial T}{\partial x}. \tag{9}$$

From this we obtain the transition jitter $\sigma_{x,Hk}$:

$$\sigma_{x,Hk} = \frac{\partial x}{\partial H_s} \sigma_{Hk}. \tag{10}$$

The corresponding values of the different media designs are shown in Table 1, where we have assumed a temperature gradient $\partial T / \partial x = 18 \mathrm{K/m}$ according to Ref(18,19) and $\sigma_{Hk} = 0.05$. Here, we have assumed that the bit transition is only defined by the thermal head field gradient. In order to keep the argument as simple as possible the magnetic head field gradient was neglected, since it is usually assumed to be much smaller than the thermal effective head field gradient. In Table 1 it is shown that indeed the transition jitter $\sigma_{x,Hk}$ which is due to variation of the anisotropy constant is very small for the FePt media and can be neglected compared to $T_c$ induced jitter as it will be shown below. The small jitter has its origin in the very high effective thermal head field gradient. The only media where $\sigma_{x,Hk}$ becomes significant is in the Fe/FePt – exchange spring media. For this media the magnetic head field gradient should also be considered for defining the bit transition, since its recording process is similar to non- heat assisted recording.

The second origin of jitter (ii) arises due to variations of the Curie temperature. Fig.3 shows the switching field for different media as function of temperature. For each media two curves are shown: one for $T_c$ and another one for $T'_c = T_c \times 1.02$. This represent a variation of $\sigma_{Tc}$ = 2%. Due to different Curie temperatures different write temperatures are required to reduce the switching field to 1.6 T. This variation is denotes with $\sigma_{Tw}$ and it is shown in Table 1. Taking into account the temperature gradient $\partial T / \partial x = 18 \mathrm{K/m}$ in the media we now can calculate the position jitter which is introduced due to the different write temperatures $T_w$ (which in turn are due to different $T_c$) as,

$$\sigma_{x,Tc} = \frac{\partial x}{\partial T} \sigma_{Tw}. \tag{11}$$

Table 1 reveals that the jitter due to variations of the Curie temperature is dominant for FePt and Fe/FePt – single phase. Interestingly this jitter is in the same order of magnitude as jitter of conventional perpendicular recording media due to variation of $H_k$.

The total transition jitter is obtained from $\sigma_{x,tot}^2 = \sigma_{x,Tc}^2 + \sigma_{x,Hk}^2$. It is worth noting that the best jitter performance can be obtained for the Fe/FePt – single domain media, although it does not has the highest effective thermal head field gradient.

To conclude a composite media is presented which leads to a very good write-ability at high temperature and a high thermal stability at room temperature. This media consists of different materials which are exchange coupled. The material on top has a Curie temperature slightly above the temperature during writing and the other material below has a significant higher Curie temperature. It is shown that for this composite media the thermally induced write error does not limit the achievable areal density in heat assisted recording. Recording can be performed at lower temperature which has besides the reduced thermal write errors several key advantages, such as an increased life time of the plasmomic transducer [20] and errors which occur after writing when the media cools down are reduced.

Furthermore, it was shown that transition jitter in heat assisted recording media is dominated by variations of the Curie temperature. Write schemes which lower the write temperature significant below the Curie temperature are beneficial in order to decrease the transition jitter although the thermal effective head field gradient becomes smaller.

The financial support of the WWTF project MA09-029 and SFB ViCoM (F4112-N13) is acknowledged.

|  | FePt | Fe/FePt– single domain | Fe/FePt–exchange spring |
|---|---|---|---|
| $T_{wr}$ [K] | 746.0 | 681.6 | 318.2 |
| $J_s (T_{wr})$ [T] | 0.166 | 1.2 | 1.75 |
| $\mu_0 dH_s/dT$ [mT/K] | 198.0 | 16.5 | 2.1 |
| $\mu_0 dH_s/dx$ [mT/nm] | 2722 | 298 | 38 |
| $\sigma_{x,Hk}$ [nm] | 0.029 | 0.26 | 2.09 |
| $\sigma_{Twr}$ [K] | 15.1 | 13.6 | 6.6 |
| $\sigma_{x,Tc}$ [nm] | 0.84 | 0.75 | 0.36 |
| $\sigma_{x,tot}$ [nm] | 0.84 | 0.79 | 2.95 |

Table 1: Write temperature ($T_{wr}$) which is required to reduce the switching field to $\mu_0 H = 1.6$ T for three different media grain designs. The average saturation polarization at this temperature is given. $\sigma_{Twr}$ shows the standard deviation of the write temperature to reduce the write field to $\mu_0 H = 1.6$ T if a standard deviation of the Curie temperature $T_c$ of $\sigma_{Tc} = 2\%$ is assumed. The resulting position jitter due to $\sigma_{Tc}$ is shown for the different media as $\sigma_{x,Tc}$. For comparison also the transition jitter $\sigma_{x,Hk}$ due to variations of the anisotropy constant is shown. $\sigma_{x,tot}$ denotes the total jitter.

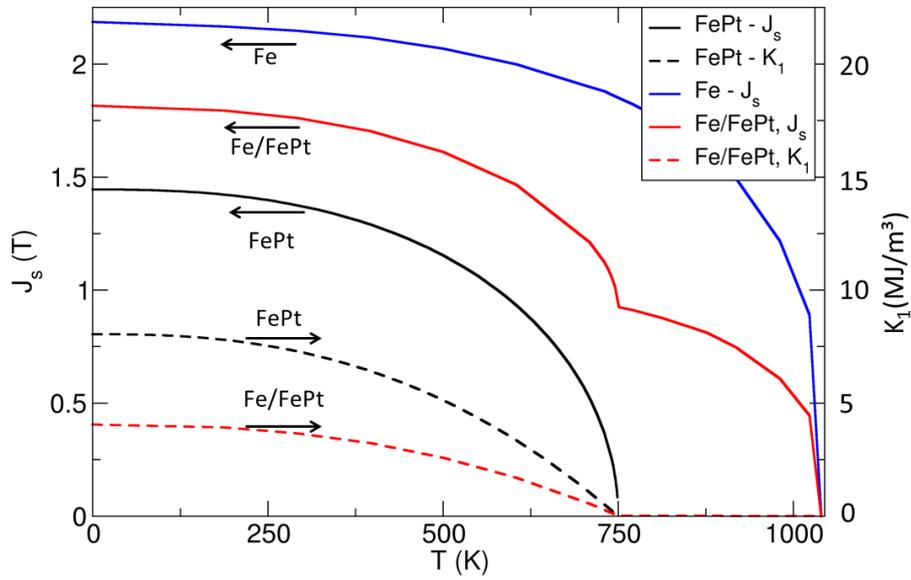

Fig. 1 (color online): Temperature dependence of saturation polarization $J_s$ and anistropy constant $K_1$ for Fe, FePt and the averaged composition Fe/FePt consisting of 50% Fe and 50% FePt.

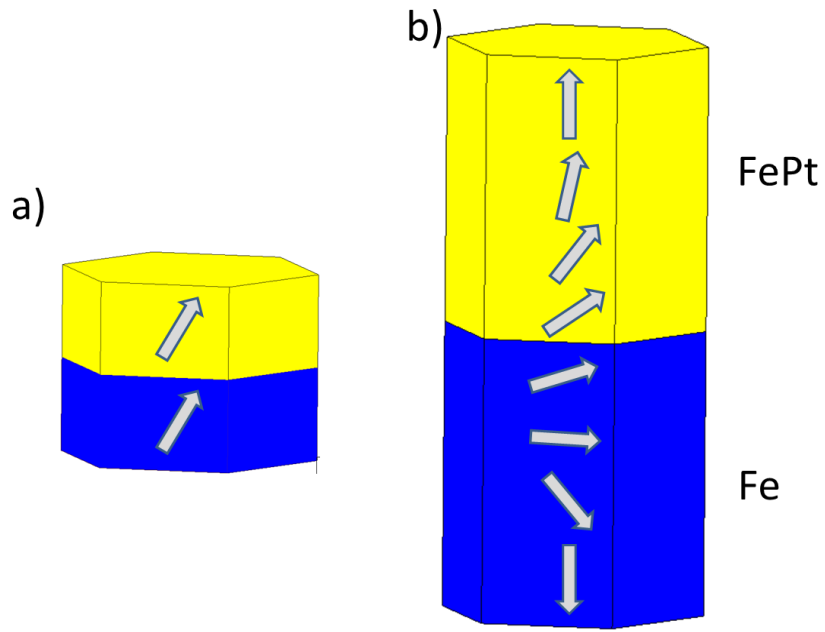

Fig. 2 (color online): Two media designs. (a: Fe/FePt - single domain) hard magnetic FePt layer is deposit on soft magnetic Fe. The column length of the grain is smaller than the exchange length of FePt. The reversal process is quasi homogenous rotation. (b: Fe/FePt - exchange spring) Column length is larger than the exchange length. The media grain reversed via formation of a domain wall.

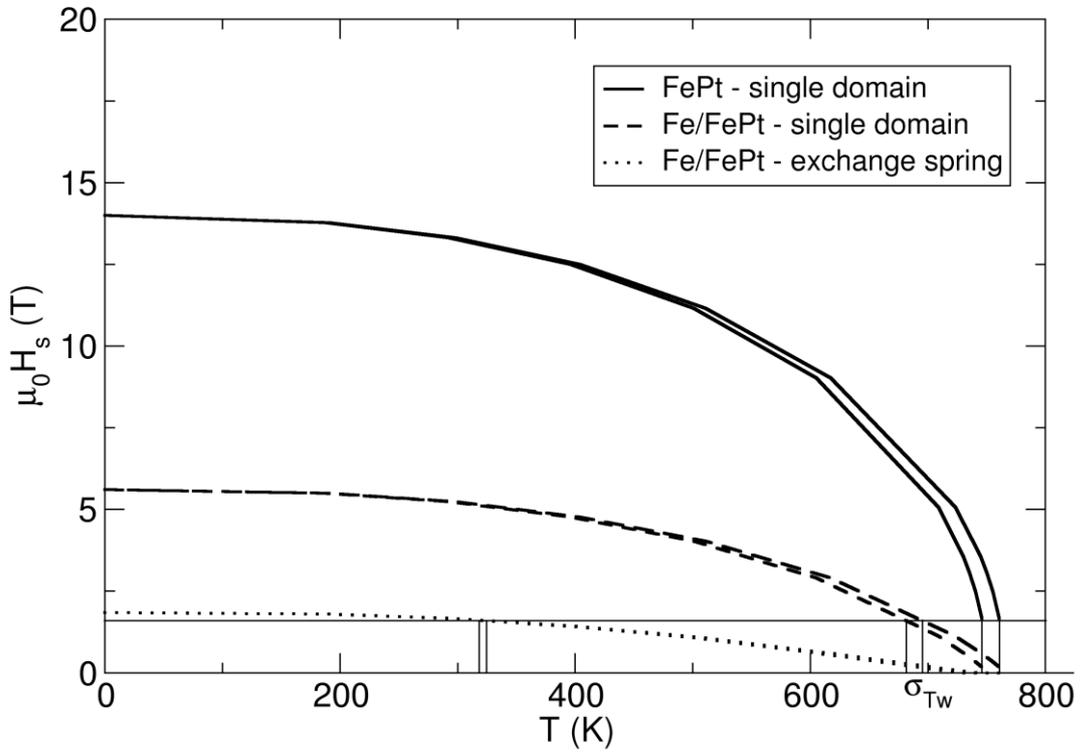

Fig. 3: Switching field of three different media grain composition as function of temperature. The second line for each switching field as function of temperature shows the plot when the Curie temperature is increased by 2%. The resulting $\sigma_T$ at which the medium switching field is reduced to 1.6 T is shown.

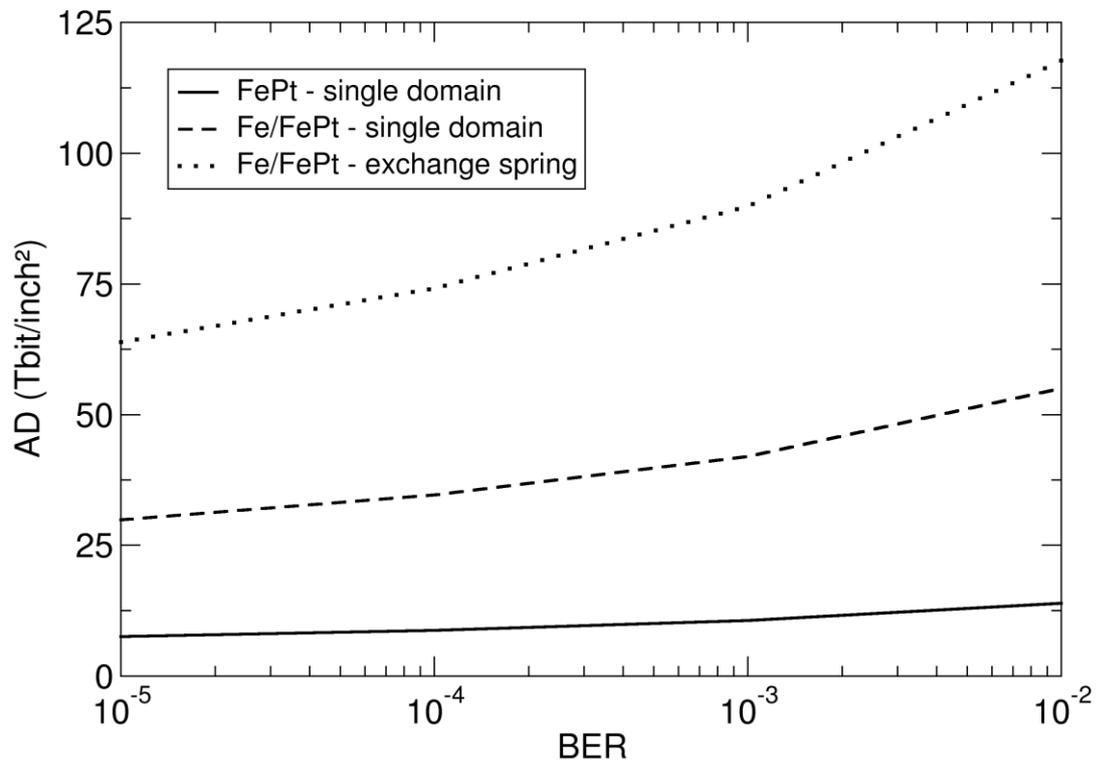

Fig. 4: Areal density (AD) of three different media grain composition as function of bit error rate (BER).